\documentclass{webofc}
\usepackage[varg]{txfonts}
\usepackage{hyperref}
\usepackage{url}
\hypersetup{colorlinks=true,citecolor=blue,urlcolor=blue,linkcolor=blue}
\usepackage{mathtools}
\usepackage{amsmath}
\usepackage{amssymb}
\usepackage{mathrsfs}
\usepackage{bm}
\usepackage{graphicx}
\usepackage{xcolor}

\begin{document}
%\preprint{BARI-TH/787-26}

\title{Probing New Physics with \texorpdfstring{$\bm{b \to u \ell \nu_\ell}$}{b -> u l nu_l} induced decays}

\author{
\firstname{Maria Letizia} \lastname{Di Cuia}\inst{1}\fnsep\thanks{\email{letizia.dicuia@clermont.in2p3.fr}} \and \firstname{Davide} \lastname{Milillo}\inst{2,3}\fnsep\thanks{\email{davide.milillo@ba.infn.it}}}

\institute{Laboratoire de Physique de Clermont Auvergne (UMR 6533), CNRS/IN2P3, Université Clermont Auvergne, 4 Av. Blaise Pascal, 63178 Aubière Cedex, France
\and
Dipartimento Interateneo di Fisica "Michelangelo Merlin", Università degli Studi di Bari, via Orabona 4, 70126 Bari, Italy
\and
Istituto Nazionale di Fisica Nucleare, Sezione di Bari, via Orabona 4, 70126 Bari, Italy}

\abstract{
We analyze $B^- \to \mu^- \bar{\nu}_\mu$, $\bar{B}^0 \to \pi^+ \mu^-\bar{\nu}_\mu$, and $B^- \to \rho^0 \mu^- \bar{\nu}_\mu$ decays using the data on binned branching fractions recently measured by the Belle II Collaboration for the semileptonic modes. 
We extend the low-energy Hamiltonian governing $b \to u \ell \bar{\nu}_\ell$ transitions to include the full set of dimension-six, Lorentz-invariant operators with left-handed neutrinos, consistent with the Standard Model Effective Field Theory (SMEFT) framework. The Wilson coefficients are treated as complex, lepton-flavor-dependent effective couplings. Since the three selected channels are sensitive to different non-standard structures, we perform a combined analysis using Monte Carlo methods to constrain the New Physics couplings for three benchmark values of $|V_{ub}|$. We find that, while the Standard Model remains broadly compatible with the data, some of the couplings show a mild preference for non-zero values \cite{DLFP}.}

\maketitle

\section{Introduction}
Semileptonic $B$ meson decays currently play a crucial role in the determination of fundamental parameters of the Standard Model (SM), namely the elements of the Cabibbo-Kobayashi-Maskawa (CKM) matrix \cite{Cabibbo:1963yz,Kobayashi:1973fv}. Moreover, as highly precise datasets become available, they provide significant tests of the SM and its potential extensions \cite{Gambino:2020jvv}.

This is indeed the case for $b \to u \ell {\bar \nu}_\ell$ induced decays: the branching fractions of both purely leptonic and semileptonic modes involve the CKM element $|V_{ub}|$, whose precise measurement represents a long-standing challenge. A tension persists between the experimental determinations of this parameter obtained using inclusive and exclusive methods, with the former yielding a larger value than the latter at the level of about $2.5 \sigma$ \cite{HeavyFlavorAveragingGroupHFLAV:2024ctg}. This discrepancy may be due to an incomplete control over hadronic form-factor uncertainties \cite{Crivellin:2014zpa}; alternatively, it may hint at New Physics (NP) effects, a scenario supported by the analogous tension in the CKM $|V_{cb}|$ matrix element \cite{Crivellin:2009sd, Colangelo:2016ymy}.

To shed light on this tension, we focus on the purely leptonic channel $B^- \to \mu^- \bar{\nu}_\mu$ and on the semileptonic modes $\bar{B}^0 \to \pi^+ \mu^-\bar{\nu}_\mu$ and $B^- \to \rho^0 \mu^- \bar{\nu}_\mu$. The recent measurements of differential branching fractions in bins of $q^2$ (the squared momentum transfer to the lepton pair) provided by the Belle II Collaboration for the semileptonic channels \cite{Belle-II:2024xwh} offer a valuable opportunity to probe non-standard interactions. The availability of binned data significantly enhances the sensitivity to them, as different NP operators affect the $q^2$ spectra in distinct ways.

In our analysis \cite{DLFP} we adopt a model-independent effective field theory approach and exploit the complementary sensitivity of the decay modes to different operator structures to constrain their effective couplings, mapping out the parameter space and studying the interplay of CKM and NP inputs. In what follows, we present a selection of our results, namely the determination of $|V_{ub}|$ within the SM and the fit of NP couplings for three benchmark values of $|V_{ub}|$ from the combined analysis of semileptonic and leptonic data.

\section{Framework}
Our analysis is based on the low-energy effective Hamiltonian
\begin{align}
\mathcal{H}_{\rm{eff}} =
\frac{G_{F}}{\sqrt{2}} &V_{ub}
\big\{ (1+\epsilon_{V}^{\ell})\bar{u}\gamma_{\mu}(1-\gamma_{5})b\bar{\ell}\gamma^{\mu}(1-\gamma_{5})\nu_{\ell} \nonumber \\[-2pt] 
&+\epsilon_{R}^{\ell} \bar{u}\gamma_{\mu}(1+\gamma_{5})b \bar{\ell}\gamma^{\mu}(1-\gamma_{5})\nu_{\ell} \nonumber\\[4pt]
&+ \epsilon_{S}^{\ell}\bar{u}b\bar{\ell}(1-\gamma_{5})\nu_{\ell} + \epsilon_{P}^{\ell}\bar{u}\gamma_{5}b\bar{\ell}(1-\gamma_{5})\nu_{\ell} \nonumber\\[4pt] &+\epsilon_{T}^{\ell} \bar{u}\sigma_{\mu\nu}(1-\gamma_{5})b \bar{\ell}\sigma^{\mu\nu}(1-\gamma_{5})\nu_{\ell} \big\} + \rm{h.c.} \label{Heff}
\end{align}
comprising the full set of dimension-six operators with left-handed neutrinos consistent with the Standard Model Effective Field Theory (SMEFT) framework \cite{Grzadkowski:2010es}. The Wilson coefficients $\epsilon_{j}^{\ell}$ are assumed as lepton-flavor-dependent complex numbers; their real and imaginary parts $\{\mathrm{Re}(\epsilon_{j}^{\ell}),\mathrm{Im}(\epsilon_{j}^{\ell})\}$ are treated as independent free parameters. The SM corresponds to $\epsilon_j^\ell=0$ for all $j$.

The observables of $B$ meson decays governed by the Hamiltonian Eq.~\eqref{Heff} exhibit distinct sensitivity to different non-standard structures. This complementarity is essential in isolating possible NP effects and constraining the coefficients \cite{Ivanov:2017hun}.

\subsection{\texorpdfstring{Leptonic $\boldsymbol{B^- \to \ell^- \bar{\nu}_\ell}$ mode}{Leptonic B- → ℓ-v̄ℓ mode}}

The decay width of the purely leptonic $B^- \to \ell^- \bar{\nu}_\ell$ mode resulting from \eqref{Heff} reads
\begin{equation}
\Gamma_\ell=\frac{G_F^2 |V_{ub}|^2f_B^2m_B^3}{8\pi }\biggl(1-\frac{m_\ell^2}{m_B^2} \biggr)^2 \left|  \left(\frac{m_\ell}{m_B} \right)  (1+\epsilon_V^\ell-\epsilon_R^\ell)+\frac{m_B}{m_b+m_u}\epsilon_P^\ell \right|^2 \,, \label{Blnu}
\end{equation}
where $f_B=190.0\pm 1.3$ MeV   \cite{FlavourLatticeAveragingGroupFLAG:2024oxs} is the $B^-$ decay constant, parametrizing the current-vacuum matrix element $\langle 0|{\bar u} \gamma_\mu \gamma_5 b |  B^- (p)\rangle= i \, f_B p_\mu$, with $p$ the $B^-$ four-momentum.

Eq.~\eqref{Blnu} shows that this mode gives access to the Wilson coefficients
\begin{equation}
 \epsilon_2^\ell=\epsilon_V^\ell-\epsilon_R^\ell \quad, \qquad \epsilon_P^\ell \quad  .
\label{epsLept}
\end{equation}
Notably, the pseudoscalar operator lifts the chiral suppression typical of the SM. The couplings are understood at the low-energy scale $\mu_b=m_b$. The running quark masses are defined in the $\overline{\mathrm{MS}}$ scheme; we take $m_b(m_b)=4.186 \pm 0.006~\mathrm{GeV}$ and $m_u(2~\mathrm{GeV})=2.16 \pm 0.07\ \mathrm{MeV}$ \cite{ParticleDataGroup:2026mpi}.

\subsection{\texorpdfstring{Semileptonic $\boldsymbol{\bar{B}^0 \to \pi^+\ell^- \bar{\nu}_\ell}$ mode}{Semileptonic B̄0 → π+ℓ-v̄ℓ}}

The $\bar{B}^0 \to \pi^+ \ell^- \bar{\nu}_{\ell}$ differential decay distribution derived from Eq.~\eqref{Heff} reads \cite{Colangelo:2019axi}
\begin{align}
\frac{d\Gamma_\ell }{dq^{2}} &= \frac{G_{F}^{2}|V_{ub}|^{2}\lambda^{1/2}}{128m_{B}^{3}\pi^{3}q^{2}}\left(1-\frac{m_{\ell}^{2}}{q^{2}}\right)^{2}
\Bigg\{ \left| m_{\ell}(1+\epsilon_{V}^{\ell}+\epsilon_{R}^{\ell}) + \frac{q^{2}\epsilon_{S}^{\ell}}{m_{b}-m_{u}} \right|^{2}(m_{B}^{2}-m_{\pi}^{2})^{2} f_{0}^2 \notag \\
&+ \frac{\lambda}{3}
\Biggl[\left| m_{\ell}(1+\epsilon_{V}^{\ell}+\epsilon_{R}^{\ell})f_{+} + \frac{4q^{2}f_{T}}{m_{B}+m_{\pi}} \epsilon_{T}^{\ell}\right|^{2}+ 2q^2 \left| (1+\epsilon_{V}^{\ell}+\epsilon_{R}^{\ell})f_{+} + \frac{4 m_{\ell}f_{T}}{m_{B}+m_{\pi}}\epsilon_{T}^{\ell} \right|^{2} \Biggr] \Bigg\},
\label{Bpilnu}
\end{align}
where $\lambda \equiv  \lambda(m_B^2,m_\pi^2,q^2)$ is the K\"all\'en function, and $q^2 \in [m_\ell^2, (m_B-m_\pi)^2]$. In Eq.~\eqref{Bpilnu}, the $\bar{B}^{0}\rightarrow\pi^{+}$ matrix elements are expressed in terms of the form factors $\left\{ f_{0}(q^2),f_{+}(q^2),f_{T}(q^2)\right\}$ adopted from Ref.~\cite{Leljak:2021vte}, where the Bourrely-Caprini-Lellouch (BCL) parametrization \cite{Bourrely:2008za} is implemented together with a suitable modification of the scalar form factor. The covariance matrix for the form factor parameters is provided in the same reference.

Eq.~\eqref{Bpilnu} shows that this mode gives access to the coefficients
\begin{equation}
\epsilon_1^\ell=\epsilon_V^\ell+\epsilon_R^\ell \quad, \qquad \epsilon_S^\ell \quad  , \qquad \epsilon_T^\ell \quad.
\label{epsPi}
\end{equation}
The pseudoscalar operator in Eq. \eqref{Heff} does not contribute due to parity conservation.

\subsection{\texorpdfstring{Semileptonic $\boldsymbol{B^- \to \rho^0\ell^- \bar{\nu}_\ell}$ mode}{Semileptonic B- → ρ0ℓ-v̄ℓ}}

The fully differential decay distribution for the semileptonic $B^- \to \rho^0(\to \pi^+ \pi^-)\ell^- \bar{\nu}_\ell$ can be found in \cite{Colangelo:2019axi,Colangelo:2024mxe}. Integrating over the angular variables, the differential decay width reads
\begin{align}
&\frac{d\Gamma_\ell}{dq^2} = \frac{G_F^2 |V_{ub}|^2}{256\pi^3 m_B^3} \left(1 - \frac{m_\ell^2}{q^2}\right)^2 \lambda^{1/2}
\Bigg\{ \frac{1}{6} \biggl( 2(H_0^{\mathrm{SM}})^2 + (H_+^{\mathrm{SM}} + H_-^{\mathrm{SM}})^2 \biggr) \, |1 + \epsilon_V^\ell-\epsilon_R^\ell|^2 \nonumber\\
\quad &+ \frac{1}{6} (H_+^{\mathrm{SM}} - H_-^{\mathrm{SM}})^2 |1 + \epsilon_V^\ell+\epsilon_R^\ell|^2 + \frac{4}{3} \left( \frac{1}{64} (H_L^{\mathrm{NP}})^2 + (H_+^{\mathrm{NP}})^2 + (H_-^{\mathrm{NP}})^2 \right) \, |\epsilon_T^\ell|^2 \nonumber\\
\quad &+ (H_t^{\mathrm{NP}})^2 \left| m_\ell^2(1 + \epsilon_V^\ell - \epsilon_R^\ell) + \frac{q^2}{m_b + m_u} \epsilon_P^\ell \right|^2 \nonumber\\
\quad &- m_\ell \sqrt{q^2} \bigg[ 2(H_+^{\mathrm{SM}} - H_-^{\mathrm{SM}})(H_+^{\mathrm{NP}} - H_-^{\mathrm{NP}}) \mathrm{Re}\bigl[\epsilon_T^\ell (1 + \epsilon_V^\ell+\epsilon_R^\ell)^*\bigr] \nonumber\\
\quad & + \frac{1}{2} \left( H_0^{\mathrm{SM}} H_L^{\mathrm{NP}} + 4(H_+^{\mathrm{SM}} + H_-^{\mathrm{SM}})(H_+^{\mathrm{NP}} + H_-^{\mathrm{NP}}) \right) \mathrm{Re}\bigl[\epsilon_T^\ell (1 + \epsilon_V^\ell-\epsilon_R^\ell)^*\bigr] \bigg] \Bigg\}.
\label{Brholnu}
\end{align}
Eq.\eqref{Brholnu} is written in terms of the helicity amplitudes $H_{0}^{\mathrm{SM}}, H_{\pm}^{\mathrm{SM}}$ and $H_{L}^{\mathrm{NP}}, \, H_{t}^{\mathrm{NP}}, H_{\pm}^{\mathrm{NP}}$, which describe the contributions of both the purely $V-A$ and NP structures to the decay. The isospin factor $1/\sqrt{2}$ between $B^0\to \rho^+$ and $B^-\to \rho^0$ has been included in Eq.~\eqref{Brholnu}. Exploiting the isospin symmetry, we adopt the Bharucha-Straub-Zwicky (BSZ) form factors, determined in Ref.\cite{Bharucha:2015bzk} via Light-Cone QCD sum rules constrained by lattice QCD data. The full correlation matrix for the form-factor parameters is provided in the same reference.
Finally, the parametrization of the helicity amplitudes in terms of the BSZ form factors is detailed in \cite{DLFP}.

As shown in Eq.~(\ref{Brholnu}), this channel provides access to the coefficients
\begin{equation}
\epsilon_{1,2}^{\ell}=\epsilon_{V}^{\ell}\pm\epsilon_{R}^{\ell} \quad , \qquad \epsilon_{P}^{\ell} \quad , \qquad  \epsilon_{T}^{\ell} \quad ,
\end{equation}
while the scalar operator does not contribute.

\section{Data and methods}
For the semileptonic modes we use the  Belle II Collaboration data collected at an integrated luminosity of $364\,\mathrm{fb}^{-1}$ \cite{Belle-II:2024xwh}. Although measurements of the decay distributions are also available from the same and other experiments 
\cite{CLEO:1999mif,Belle:2006hlt,CLEO:2007vpk,Belle:2010hep,BaBar:2010efp,BaBar:2012thb,Belle:2013hlo}, we restrict our analysis to this dataset only to ensure a consistent treatment of the systematic uncertainties across the two modes. The provided data represent the average of the electron and muon samples. Assuming lepton-flavor universality for the light states ($\epsilon_j^e=\epsilon_j^\mu$), we interpret the combined $e/\mu$ measurement as a constraint on the corresponding muon coefficients.

The partial branching ratios for the $\bar{B}^0 \to \pi^+ \mu^- \bar{\nu}_\mu$ and $B^- \to \rho^0 \mu^- \bar{\nu}_\mu$ modes,
\begin{equation}
    \Delta \mathcal{B}_{i}^{\rm exp} =
    \int_{q_i^2}^{q_{i+1}^2}
    dq^2 \, \frac{d\mathcal{B}}{dq^2} \ ,
\end{equation}
cover the full kinematic range by 13 and 10 bins of $q^2$, respectively. The corresponding binned theoretical branching fractions are obtained integrating the differential decay distributions defined in Sec.~2,
\begin{equation}
    \Delta \mathcal{B}_{i}^{\rm th}
    (|V_{ub}|, \epsilon_j^\mu)=
    \int_{q_i^2}^{q_{i+1}^2}
    dq^2 \, \frac{d\mathcal{B}}{dq^2}(q^2, |V_{ub}|, \epsilon_j^\mu) \ .
\end{equation}

The uncertainties are propagated through the full covariance matrix,
\begin{equation}
    C = C_{\rm exp} + C_{\rm th}.
\end{equation}
Specifically, the experimental covariance matrix $C_{\rm exp}$ is constructed combining statistical and systematic errors on binned branching fractions provided in Ref.~\cite{Belle-II:2024xwh}. On the other hand, the theoretical covariance matrix $C_{\rm th}$ is obtained by a linearized propagation of the form-factor uncertainties, and accounts for bin-to-bin correlations through its off-diagonal terms. The construction of $C_{\rm th}$ is detailed in Ref.~\cite{DLFP}.

To constrain the free parameters $\boldsymbol{\theta}_j=\{\mathrm{Re}(\epsilon_j^\mu), \mathrm{Im}(\epsilon_j^\mu)\}$ $(j=1,2,S,P,T)$, we apply a Markov Chain Monte Carlo (MCMC) analysis based on the Metropolis-Hastings algorithm \cite{Trotta_2008,ParticleDataGroup:2026mpi,10.1214/ss/1015346320}. The starting point is the construction of the log-likelihood function,
\begin{equation}
    \log \mathcal{L}= -\frac{1}{2}
    \bigl\{\chi^2 + \log[\det(C)]\bigr\} + \mathrm{const}.
\label{logL}
\end{equation}
The $\chi^2$ function is given by
\begin{equation}
    \chi^2 = (\Delta \mathcal{B}_{\mathrm{exp}}-\Delta \mathcal{B}_{\mathrm{th}})^T \, C^{-1}
    (\Delta \mathcal{B}_{\mathrm{exp}}-\Delta \mathcal{B}_{\mathrm{th}}),
\label{chi2}
\end{equation}
where $(\Delta \mathcal{B}_{\mathrm{exp}}-\Delta \mathcal{B}_{\mathrm{th}})$ is the column vector of residuals, spanning the bins of each channel. Since $C_{\rm th}$ depends on NP couplings, the $\log[\det(C)]$ term cannot be incorporated into the additive constant; this requires recomputing $C_{\rm th}$ at each step of the Markov chain in parameter space. For the same reason, the inversion of the full covariance matrix in Eq.\eqref{chi2} must be performed numerically.

By running 4 chains of $5\times 10^5$ steps each starting from random points within the chosen prior intervals, we sample the marginal posterior distribution $p(\boldsymbol{\theta}_j|\mathrm{data})$ and determine the highest posterior density (HPD) regions for the free parameters. A $(1-\alpha)\times100\%$ HPD region is the smallest region of parameter space having the $(1-\alpha)$ probability of containing the true value of the parameters,
\begin{equation}
    \mathcal{R}_{\alpha}^{(j)}=\bigl\{
    \boldsymbol{\theta}_j \ \bigl| \ p(\boldsymbol{\theta}_j|\mathrm{data})\geq p_{\alpha}
    \bigr\},
\label{hpd}
\end{equation}
where the threshold $p_{\alpha}$ is the largest number such that
\begin{equation}
    \iint_{\mathcal{R}_{\alpha}^{(j)}}
    p(\boldsymbol{\theta}_j |\mathrm{data}) \, d^2\boldsymbol{\theta}_j=1-\alpha.
\label{alphaint}
\end{equation}
The target distribution $p(\boldsymbol{\theta}_j |\mathrm{data})$ is numerically reconstructed from the whole set of sampled points via kernel density estimation (KDE). Technical details on the implementation of MCMC, choice of priors, and diagnostics can be found in Ref.\cite{DLFP}.

The results obtained with the MCMC method have been independently cross-checked through a profile-likelihood analysis, accelerated by a supervised classifier based on the Random Forest algorithm. The detailed implementation of this procedure is illustrated in Ref.~\cite{DLFP}. The two methods identify similar regions in parameter space, with minor expected differences arising from the distinct statistical frameworks on which they rely.

\section{Analysis and selected results}
\subsection{\texorpdfstring{Fit of $|V_{ub}|$ in the SM}{Fit of |Vub| in the SM}}
Assuming the SM, we first determine $|V_{ub}|$ combining the data on semileptonic modes reported by Belle II Collaboration with the measurement of the branching fraction of the purely leptonic mode $B^- \to \tau^- \bar{\nu}_\tau$ \cite{ParticleDataGroup:2026mpi}. We find
\begin{equation}
    |V_{ub}|_{\rm SM}=
    (3.91 \pm 0.17)\times 10^{-3}\, .
\end{equation}
a result compatible with other determinations \cite{ParticleDataGroup:2026mpi, HeavyFlavorAveragingGroupHFLAV:2024ctg}.

\subsection{Global fit of NP couplings}
To investigate the separate effect of $|V_{ub}|$ and NP couplings, we choose three benchmark values $|V_{ub}|_{\min}, \, |V_{ub}|_{\rm mid},\, |V_{ub}|_{\max}$ and determine the two-dimensional HPD regions for the $\{\mathrm{Re}(\epsilon_j^\mu), \mathrm{Im}(\epsilon_j^\mu)\}$ pairs. Specifically, 
$|V_{ub}|_{\min}$ corresponds to the exclusive determination, while $|V_{ub}|_{\max}$ lies in the highest range of the measurements from inclusive  semileptonic $B$ decays \cite{HeavyFlavorAveragingGroupHFLAV:2024ctg}; finally $|V_{ub}|_{\rm mid}$ is the average between the two.

We run our analysis for individual modes, $B^- \to \mu^- \bar{\nu}_\mu$ (for which only an experimental upper limit is available), $\bar{B}^0 \to \pi^+ \mu^-\bar{\nu}_\mu$, and $B^- \to \rho^0 \mu^-\bar{\nu}_\mu$. Regarding the latter, having verified that the $B^- \to \mu^- \bar{\nu}_\mu$ mode imposes a much stronger bound (by about one order of magnitude) on $\epsilon_P^\mu$, we restrict the prior on $\mathrm{Re}(\epsilon_P^\mu)$ and $\mathrm{Im}(\epsilon_P^\mu)$ from the start to account for the information from the leptonic channel. The decays are then combined into a global fit. Here we describe only the latter step; the results obtained for individual modes are also instructive and can be found in Ref. \cite{DLFP}.

\begin{figure*}[t]
\centering
\includegraphics[width = 0.19\linewidth]{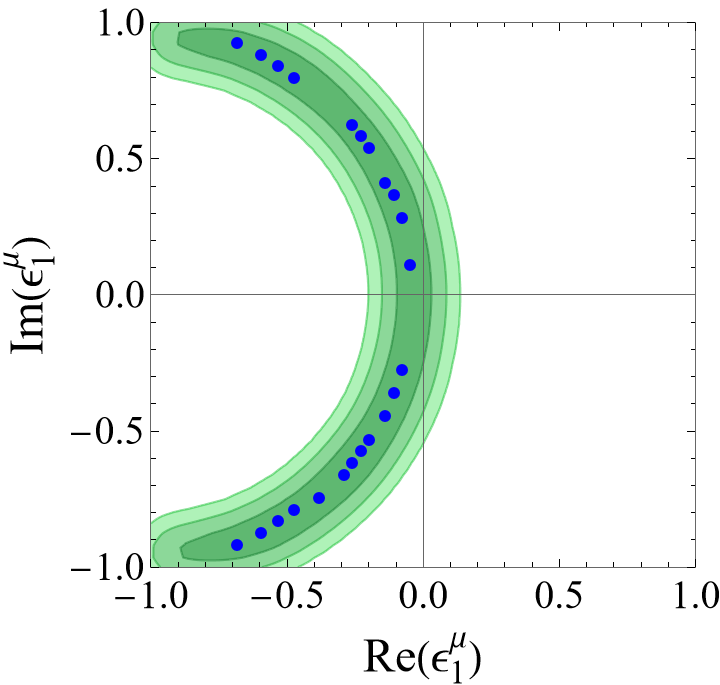} 	
\includegraphics[width = 0.19\linewidth]{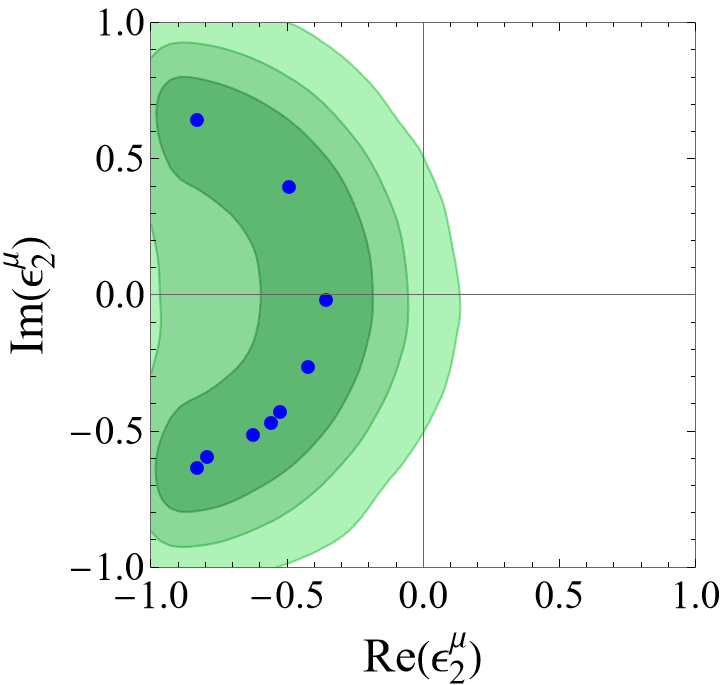}
\includegraphics[width = 0.19\linewidth]{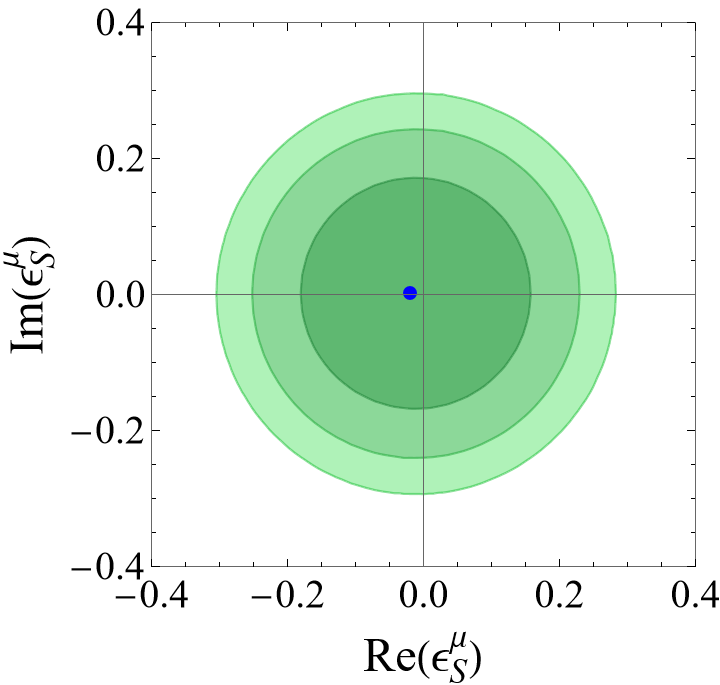}
\includegraphics[width = 0.19\linewidth]{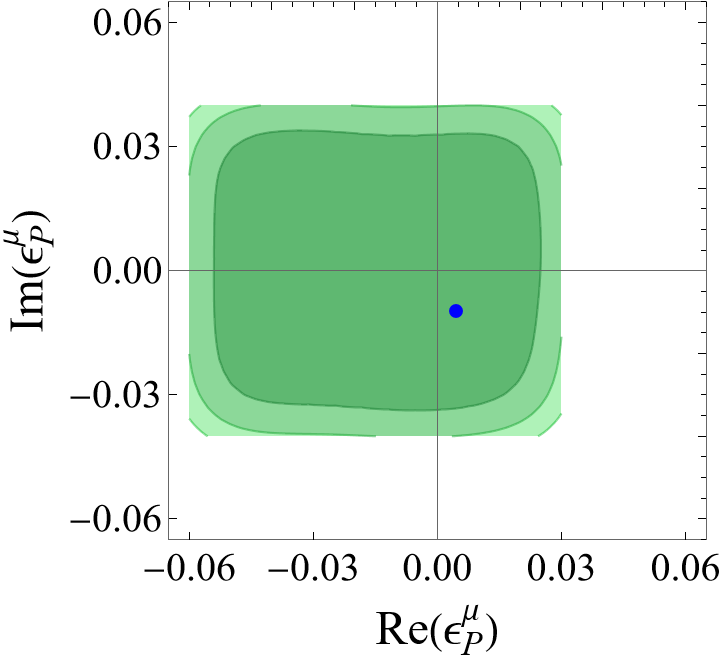}
\includegraphics[width = 0.19\linewidth]{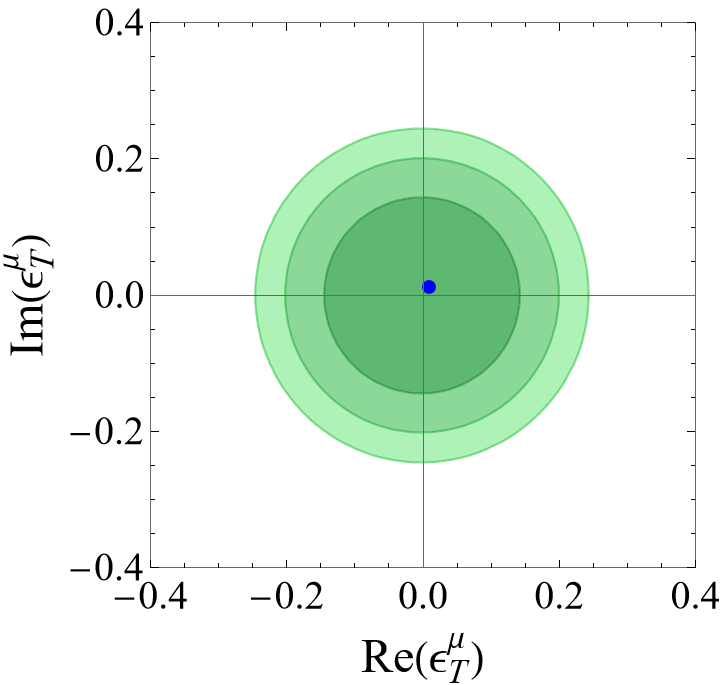}\\
\includegraphics[width = 0.19\linewidth]{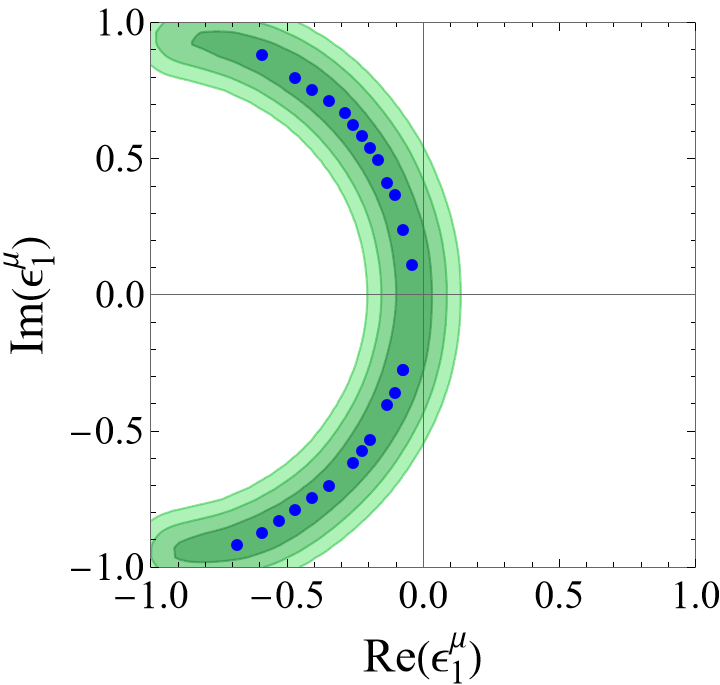} 	
\includegraphics[width = 0.19\linewidth]{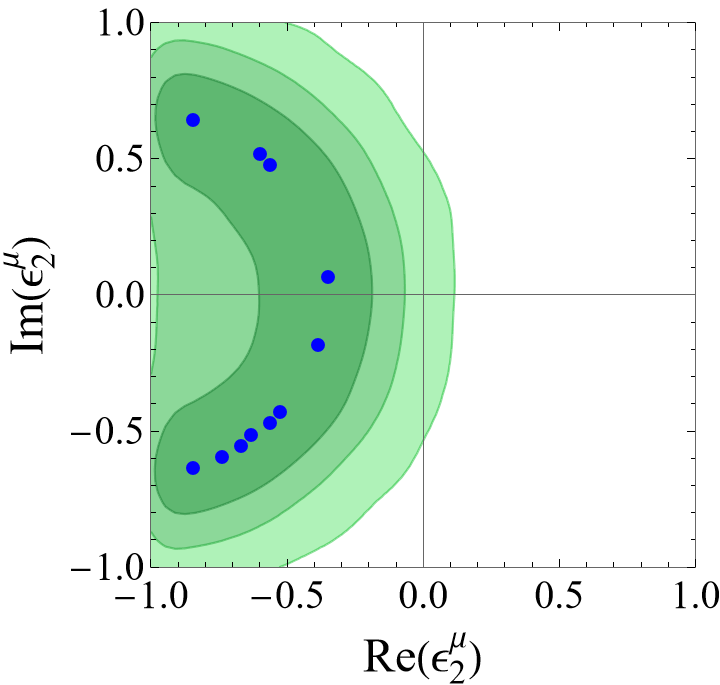}
\includegraphics[width = 0.19\linewidth]{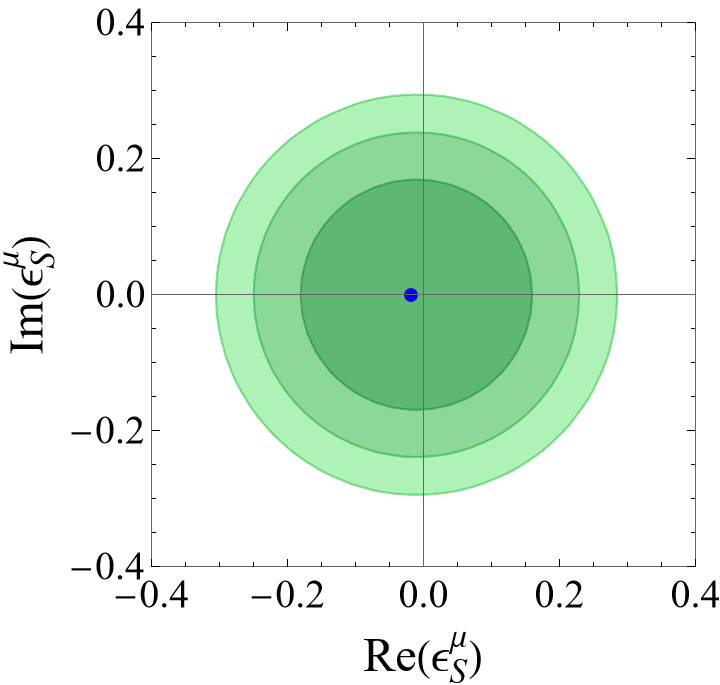}
\includegraphics[width = 0.19\linewidth]{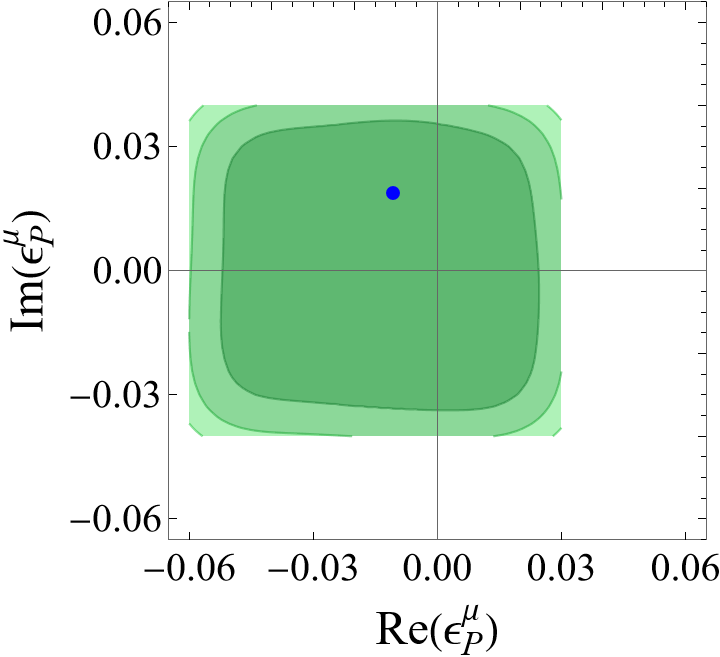}
\includegraphics[width = 0.19\linewidth]{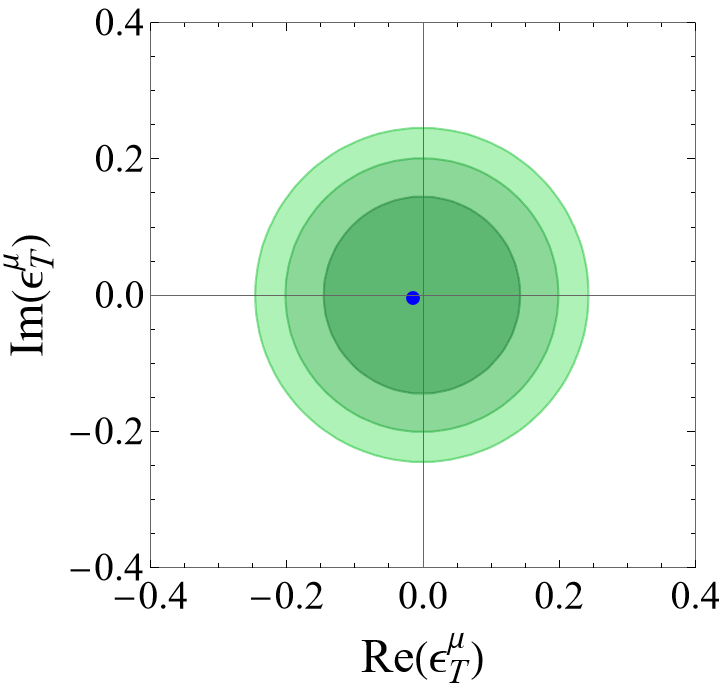}\\
\includegraphics[width = 0.19\linewidth]{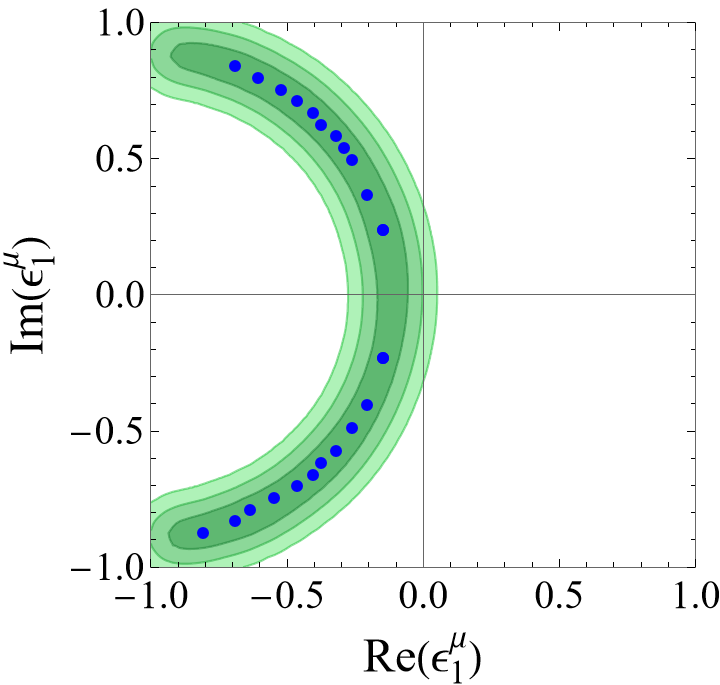} 	
\includegraphics[width = 0.19\linewidth]{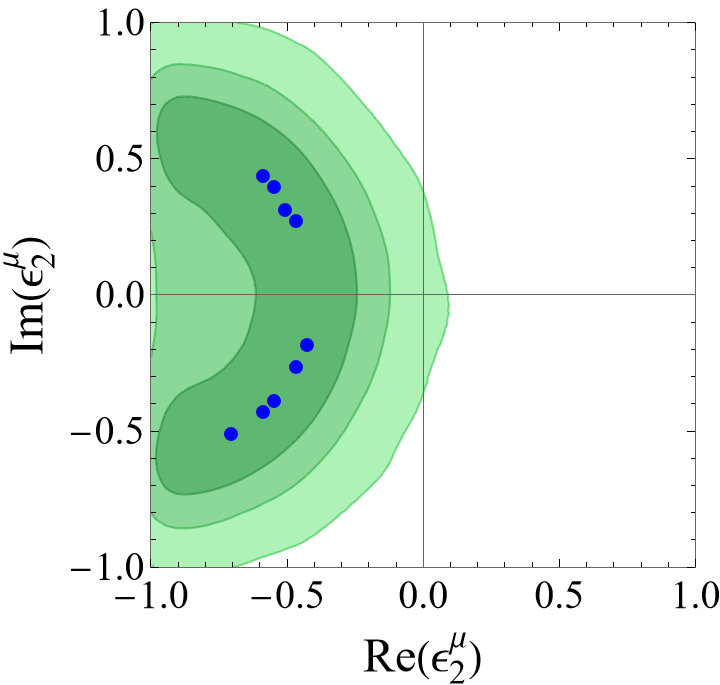}
\includegraphics[width = 0.19\linewidth]{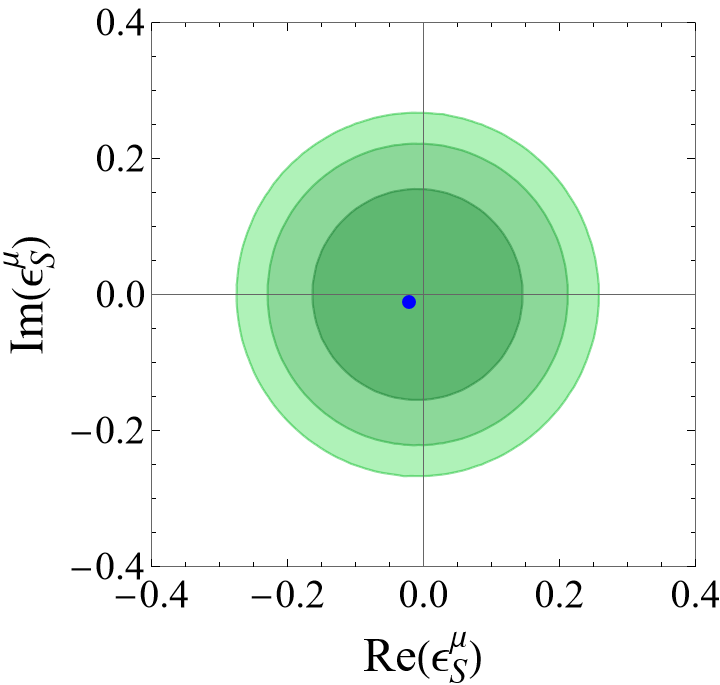}
\includegraphics[width = 0.19\linewidth]{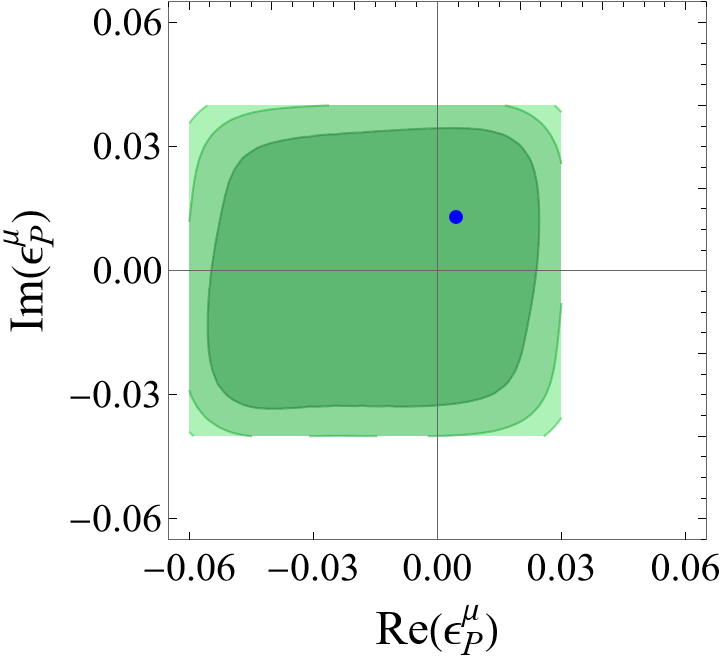}
\includegraphics[width = 0.19\linewidth]{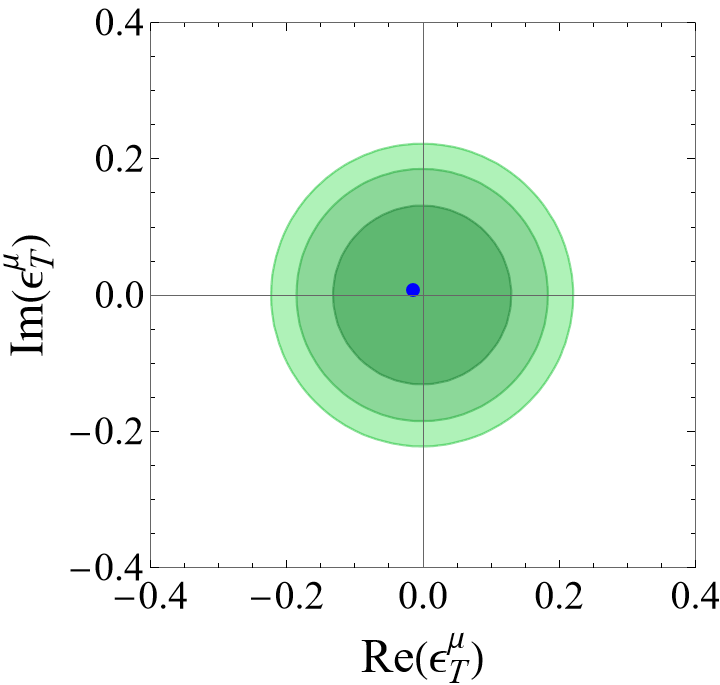}
\caption{
Allowed regions of parameter space for $\epsilon_{1,2}^\mu=\epsilon_V^\mu \pm \epsilon_R^\mu$, $\epsilon_S^\mu$, $\epsilon_P^\mu$ and $\epsilon_T^\mu$ obtained from the combined analysis of the modes $B^- \to  \mu^- \bar \nu_\mu$, $\bar B^0 \to \pi^+ \mu^- \bar \nu_\mu$ and $B^- \to \rho^0 \mu^- \bar \nu_\mu$. From top to bottom, the plots refer to the benchmark values $|V_{ub}|_{\mathrm{min}}$,  $|V_{ub}|_{\mathrm{mid}}$, $|V_{ub}|_{\mathrm{max}}$, respectively. Shaded regions, from darkest to lightest, represent the $68.3\%, \, 95.4\%$ and $99.7\%$ HPD regions. The dark dots represent the global maxima of the posterior distribution; see the text for the discussion in the case of $\epsilon_1^\mu$ and $\epsilon_2^\mu$.}
\label{figAll}
\end{figure*}

Fig.[\ref{figAll}] displays the $68.3\%, \, 95.4\%$ and $99.7\%$ HPD regions in the $\left(\mathrm{Re}(\epsilon_j^\mu), \mathrm{Im}(\epsilon_j^\mu)\right)$ complex plane. For some of the couplings, namely $\epsilon_{1}^\mu$ and $\epsilon_{2}^\mu$, the topology of the sampled regions suggests the presence of a flat direction in parameter space. The origin of this feature can be understood from the structure of the decay amplitudes. The vector and axial contributions involve the combinations $1+\epsilon^\mu_{1,2}$, so that the dominant quadratic terms in the decay distributions depend on
\begin{equation}
|1+\epsilon^\mu_{1,2}|^2=
(1+\mathrm{Re}\,\epsilon^\mu_{1,2})^2+
(\mathrm{Im}\,\epsilon^\mu_{1,2})^2 \,\,.
\end{equation}
Therefore, the data mainly constrain the moduli $|1+\epsilon_{1,2}^\mu|$ rather than the real and imaginary parts separately: configurations with the same $|1+\epsilon^\mu_{1,2}|$ contribute identically\footnote{A similar effect is described in \cite{Murgui:2019czp}.} and lie on a circle centered at $(-1,0)$, truncated by the prior $|\mathrm{Re}(\epsilon_{1,2}^\mu)| \leq 1$ to the semicircle displayed in Fig.[~\ref{figAll}], where the likelihood is nearly flat.

To verify this, we perform numerically a continuous search for local maxima of the sampled posterior density. We find that the probability distributions for $\epsilon_1^\mu$ and $\epsilon_2^\mu$ possess nearly degenerate maxima distributed along an arc centered at $(-1,0)$, with small fluctuations compatible with the finite resolution of the Monte Carlo sampling and of the kernel density estimation, an indication that the local maxima are in fact statistically equivalent. Hence, the current data cannot resolve the degeneracy of the marginal posterior distributions of $\epsilon_1^\mu$ and $\epsilon_2^\mu$, and no single best-fit point can be identified; this is not the case for $\epsilon_S^\mu$, $\epsilon_P^\mu$, $\epsilon_T^\mu$, which all have well-isolated global maxima. This is illustrated in Fig.[\ref{figAll}], in which the position of the maximum probability is indicated by dark dots.

A notable result of the combined analysis is that ${\rm Re}(\epsilon_2^\mu)=0$ is disfavored at the $95.4\%$ level for all values of $|V_{ub}|$ and that ${\rm Re}(\epsilon_1^\mu)=0$ is disfavored at the $68.3\%$ level for $|V_{ub}|_{\rm max}$. For both, the significance level is understood to be local, i.e. not corrected for a look-elsewhere effect. These results reflect the information carried by individual semileptonic data samples. In the SM, where $|V_{ub}|$ is the only free parameter, the individual fit of the $B^- \to \rho^0 \mu^-\bar{\nu}_\mu$ channel alone favors a value of $|V_{ub}|$ lower than $|V_{ub}|_{\rm min}$ in order to reproduce the measured branching fractions, as found by both the Belle II analysis and our own. Conversely, if $|V_{ub}|$ is fixed at a benchmark value and NP couplings are added, the same information is transferred mainly to $\epsilon_2^\mu$ (which is dominant for the $B^-\to \rho^0$ channel), and the fit favors non-vanishing values of $\mathrm{Re}(\epsilon_2^\mu)$. A similar effect is observed, to a lesser extent, for the fit of the $\bar{B}^0 \to \pi^+ \mu^-\bar{\nu}_\mu$ mode alone, where $\mathrm{Re}(\epsilon_1^\mu)=0$ is excluded at the $68.3\%$ level for $|V_{ub}|_{\rm max}$.

These results show that a lower $|V_{ub}|$ and non-vanishing NP couplings cannot yet be disentangled, especially for the axial contribution: our analysis singles out a range of $\epsilon_2^\mu$ values that can be tested by measuring other observables sensitive to this coupling, such as the fully differential distribution of $B^-\to\rho^0(\pi^+\pi^-)\mu^-\bar\nu_\mu$ and other $b\to u\mu\bar\nu_\mu$ modes sensitive to the axial current. These measurements, combined with an improved control over hadronic uncertainties, will significantly enhance the sensitivity to potential NP effects.

\section{Conclusions}
We have presented selected results from the analysis conducted in Ref.\cite{DLFP}, which investigates potential NP effects in $b\to u\mu^-\bar{\nu}_\mu$ induced decays with a model-independent approach. The low energy Hamiltonian \eqref{Heff} governing these modes is expanded to include the full set of dimension-six operators with left-handed neutrinos, featuring lepton-flavor-dependent, complex Wilson coefficients, whose real and imaginary parts are treated as free parameters.

To constrain these parameters, we used the binned $q^2$ distributions for the semileptonic $\bar{B}^0 \to \pi^+ \mu^- \bar{\nu}_\mu$ and $B^- \to \rho^0 \mu^- \bar{\nu}_\mu$ modes reported by the Belle II Collaboration, alongside the most recent data for the purely leptonic $B^- \to \ell^- \bar{\nu}_\ell$ modes ($\ell=\mu, \tau$). The fitting procedure was conducted via MCMC and independently cross-checked by a profile-likelihood analysis using machine learning techniques.

First, we determined the CKM element $|V_{ub}|$ within the SM; here, the assumption of lepton-flavor universality allowed us to combine the semileptonic muon channels with the tauonic mode $B^- \to \tau^- \bar{\nu}_\tau$. Then, introducing the effective NP couplings $\epsilon_j^\mu$, we focused on the muonic channels $\bar{B}^0 \to \pi^+ \mu^- \bar{\nu}_\mu$ and $B^- \to \rho^0 \mu^- \bar{\nu}_\mu$, applying the constraint from the experimental upper limit on the $B^- \to \mu^- \bar{\nu}_\mu$ branching fraction. These three decays provide complementary information regarding the different operator structures, and their joint analysis yields stronger constraints than those obtained from any individual mode.

We have determined the two-dimensional HPD regions in the $\left(\mathrm{Re}(\epsilon_j^\mu), \mathrm{Im}(\epsilon_j^\mu)\right)$ complex plane for three benchmark values of $|V_{ub}|$. The global fit disfavors $\mathrm{Re}(\epsilon_2^\mu)= 0$ at the (local) $95.4\%$ level for all three benchmarks, and $\mathrm{Re}(\epsilon_1^\mu)=0$ at the (local) $68.3\%$ level for the largest $|V_{ub}|$ benchmark. Although these findings cannot be interpreted as evidence for NP, they highlight specific regions of parameter space to be further investigated as more precise measurements become available.

\section*{Acknowledgements}
We thank the coauthors of Ref.\cite{DLFP}. This work has been carried out within the INFN project (Iniziativa Specifica) SPIF.

\bibliography{ref}

\end{document}